 \documentclass[12pt,preprint]{aastex}

\shorttitle{The Galactic Anticenter Stellar Stream}

\shortauthors{Rocha-Pinto, Majewski, Skrustkie \& Crane}

\begin{document}


\title{Tracing the Galactic Anticenter Stellar Stream with 2MASS M Giants}

\author{Helio J. Rocha-Pinto, Steven R. Majewski, M. F. Skrutskie, Jeffrey D. Crane} 
\affil{Department of Astronomy, University of Virginia,
    Charlottesville, VA 22903}
\email{helio, srm4n, mfs4n, jdc2k@virginia.edu}

\begin{abstract}

The recently discovered, ring-like structure just outside the 
Galactic disk in Monoceros is detected and traced among 2MASS M giant
stars.  We have 
developed a method to recover the signature of this structure from 
the distance probability density function of stars along a 
given line of sight.  
Its detection is 
possible even when the metallicity is unknown, provided that the 
structure is not too embedded in the disk.  Application 
of this method reveals the presence of a large 
group of M giant stars at a Galactocentric distance of 18$\pm 2$ kpc, over
$+36\degr<b<+12\degr$ and $100\degr < l <270\degr$. 
Evidence that the stream extends to high negative 
latitudes is also found.   
That the structure contains M giants shows that
it contains populations of at least an order of
magnitude higher abundance than the [Fe/H] = $-1.6$ mean metallicity 
previously reported for this system.  
The structural characteristics of the stellar stream as 
traced by M giants do not support the 
interpretation of this structure as a homogeneously dense ring that 
surrounds the Galaxy, but as a more localized structure, possibly a merging 
dwarf galaxy with tidal arms, like the Sagittarius dwarf galaxy. 

\end{abstract}


\section{Introduction}

The 2MASS database includes the first homogeneous, all-sky photometric 
survey of Milky Way stars and has already reshaped our understanding of 
the structure of our Galaxy and its satellite system
(e.g., Skrutskie et al. 2001; Weinberg \& Nikolaev 2001; 
Majewski et al. 2003, ``M03'' hereafter).
A particular advantage of the NIR photometry in 2MASS is its ability  
to probe highly dust-obscured regions \citep{dutrabica,ivanovetal,dutra}. 
Thus, 2MASS has the potential to provide new insight into the structure of
the low-latitude stellar feature recently reported by Newberg et al.
(2002, hereafter N02) near the Galactic anticenter in Monoceros, and
further studied with optical analyses by Yanny et al. (2003) and 
Ibata et al. (2003; hereafter ``Y03'' and ``I03'', respectively).  These 
previous studies conclude that this highly obscured, low latitude structure 
spans a large swath across the sky, but is narrowly distributed in distance.   
I03 showed that the feature presents a very narrow main sequence, and, from
Sloan Digital Sky Survey spectra of some apparent turn-off
stars, Y03 conclude this population has [Fe/H] = $-1.6 \pm 0.3$.  

However, in their study of 2MASS-selected M giants 
with $J-K_s \ge 1.0$, M03 show evidence (see, e.g., their Figure 14)
for the Monoceros (``Mon") feature at 
the same Galactic locations as found by N02.  
If the Mon structure contains a substantial population of 
M giants, then it must contain
stars substantially more enriched than [Fe/H] = $-1.6$.  In dereddened $K_s$
magnitudes the Mon feature is a fluffy structure similar to that 
shown by N02 (see Figure 19 in M03), but
is a more coherent, narrower feature when $K_s$ 
magnitudes are converted to distances after assuming a color-magnitude 
relation (CMR) appropriate to
the Sgr core, which is dominated by an [Fe/H] $\sim -0.4$ population. 
This provides circumstantial evidence for the presence of a Mon
population with an approximately similar mean [Fe/H] 
(at least for that population 
contributing M giants), because adopting a red giant CMR 
appropriate to a substantially different metallicity \citep{ivanov} 
would presumably translate to an inflated distance spread.  

We explore the Mon 2MASS M giant density enhancement
with a technique that isolates coherent spatial features of uncertain metallicity.  
This also allows us to  
increase the statistical sample of giants over that in M03 by probing to a lower RGB 
luminosity limit (bluer $J-K_s$ color), and thereby trace the Mon feature over
230 degrees in length.

\section{Data Selection and density functions }

The NIR two-color diagram allows ready discrimination of M dwarfs and giants
(Bessell \& Brett 1988).  Here we 
employ the same 2MASS M giant two-color selection
used by M03, but expand our color range to include stars
with $0.86 < J-K_S \le 1.10$.  This provides sufficient sample 
sizes to allow probes of individual $4\degr\times 4\degr$ fields centered to $|b| > 34\degr$.
We have excluded photometry from the
51 ms 2MASS exposures, which creates a magnitude limit of $K_s \gtrsim 8$ in
our catalogue.  The typical number of M giants in fields at each
latitude range are 14-27 for $|b|\approx 34\degr$ and 51-395 for 
$|b|\approx 14\degr$.
Because our primary goal is
to {\it detect} the presence of the Mon system and determine its
length and orientation, analysis is limited to fields only
mildly ($E_{B-V} < 0.555$) reddened and minimally affected by
bulge contamination; thus regions within $|b| < 12\degr$ and $-54\degr < l  < +54\degr$
are avoided.

Since we do not know the [Fe/H] of each M giant in 
our sample, their absolute magnitudes are uncertain.  
For ascertaining relative distances (and assuming a nearly monometallic 
population) we could work with very narrow color range samples to 
minimize the dispersion in derived distance modulus.  However, 
such samples would have rather low M giant density, and larger 
sky areas would be needed to recover sample statistics, reducing our angular
resolution.

Rather than make histograms of assumed single distance values, 
we use an alternative, more robust method to find structures within the 
database via a generalized representation for the stellar distance distribution: 
Each star distance is
given by a probability density function with shape determined by the 
metallicity uncertainty of the star.   The metallicity probability density 
function (``MPDF"), $P({\rm [Fe/H]})$, gives rise to a probability density function  
for the absolute magnitude ($M_{K_S}$) of each star, which, in turn, will be 
reflected in a distance 
probability density function (``DPDF").  From the transformation 
law of probabilities \citep{press}, then
\begin{equation}
P(D\vert K_S,J-K_S) = P({\rm [Fe/H]})\cdot \left\vert {{\rm d [Fe/H]}\over 
{\rm d}M_{K_S}} \right\vert \cdot 
\left\vert {{\rm d}M_{K_S} \over {\rm d}D} \right\vert,
\label{pdfeq}
\end{equation}
where $P(D\vert K_S,J-K_S)$, hereafter $P(D)$, is the DPDF for a given apparent magnitude $K_S$ 
and $J-K_S$. 
The MPDF is certainly a complex function that 
depends on the line of sight.  For simplicity,  
we approximate it by a single-peaked function. 
By doing this, we expect more easily to find structures having 
narrower metallicity and distance distributions than Galactic disk 
stars, which present broad distributions. The MPDF peak metallicity affects the 
scale of the distance distribution, but it will not destroy any preexisting 
stellar distance groupings. 

We have adopted the metallicity-dependent CMR 
for K-M giants given by \citet{ivanov}.  The MPDF was modeled as 
a Gaussian with $\mu = -1.0$ dex and $\sigma = 0.4$ dex, 
chosen as a compromise between the contribution by
a relatively metal-rich population that we suspect exists in the structure and the
 $-1.6\pm 0.3$ dex metallicity quoted by Y03. 
The calculation of $P(D)$ is done as follows. Given $J$ and $K_S$, each [Fe/H]
corresponds to a unique $D$, which is used to substitute the metallicity in 
the right side of Eq. 
\ref{pdfeq}. This is done by using \citet{ivanov}'s [Fe/H]$(M_{K_S})$ relation 
and the distance modulus equation.  
The derivatives in Eq. \ref{pdfeq} are also calculated from these equations. 
The individual 
DPDFs for each star are then summed to yield the density function along a given line 
of sight.
The strength of the method is that by focusing 
on the DPDF, we can isolate the relative distance range with the highest expectation of 
stars.

\begin{figure}
\epsscale{0.85}
\plotone{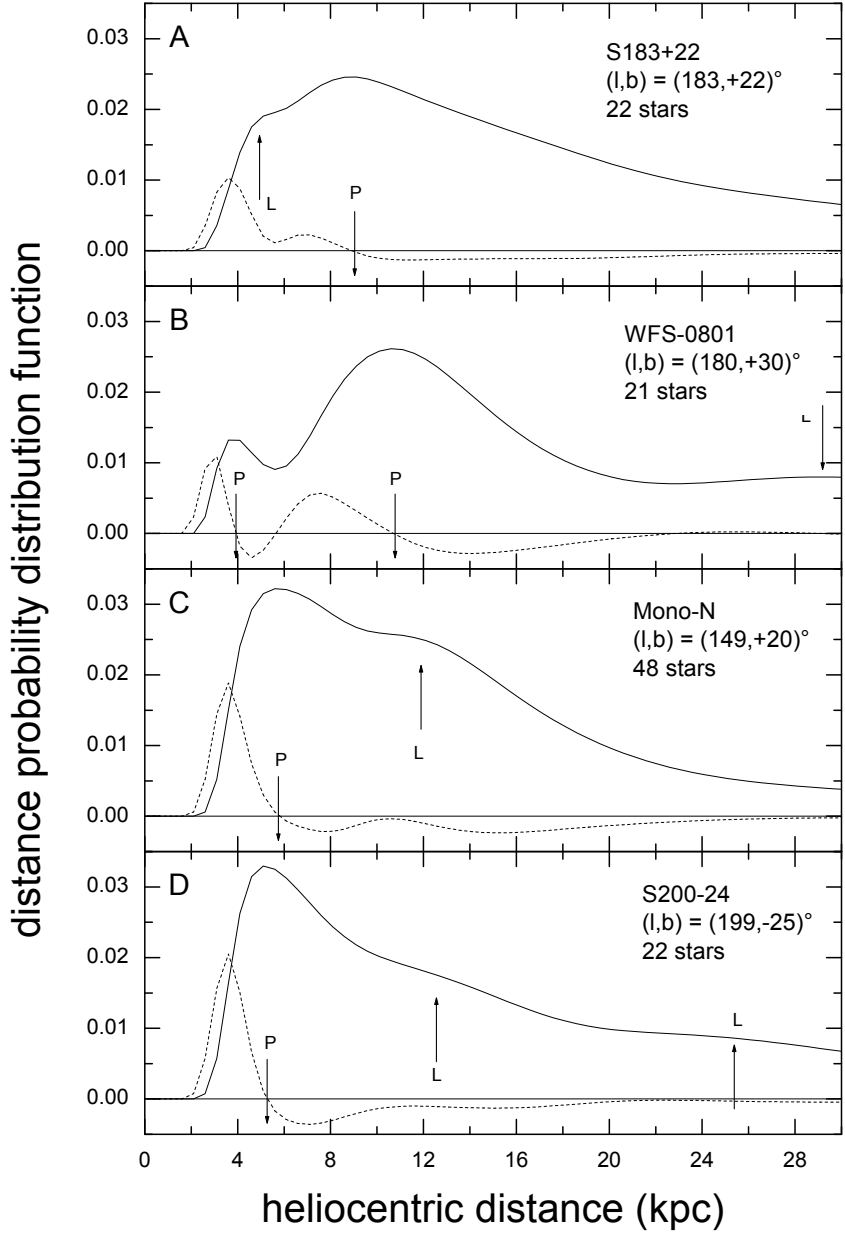}
\caption{Definition of peaks and ledges in the DPDF for lines of sight 
centered at four stellar fields where the Mon 
structure has previously been found by I03 or Y03.  A peak (labelled P) is a 
maximum value of the DPDF (global or local).  Examples of ledges (labelled L)  are also 
shown, althought they are not used in the analysis.  The first derivative 
of the DPDF (dashed lines) is used to find the position of the peaks. 
The presence of a peak (or ledge) 
between 9 and 13 kpc corresponds to our detection of the Mon structure in those fields.
\label{ledgepeak}}
\end{figure}

We are interested in discovering at which distances the DPDF peaks
for each stellar field.  The examples in Fig.~\ref{ledgepeak} demonstrate
the general character of the DPDF: In general, a broad
feature, corresponding to the disk, is found at nearby distances
(that it peaks at typically 5-7.5 kpc is the result of
missing bright M giants in our adopted 2MASS sample).
In some lines of sight, a second farther peak (sometimes stronger than the first) 
is found between 9 and 13 kpc from the Sun. For other lines of sight, an underlying 
peak blends into the broader distribution, creating a ``ledge".  
These ledges also mark preferred stellar distances in the DPDF. 
Derivatives of the DPDF locate the position of peaks (Fig.  \ref{ledgepeak}) as  
global or local maxima of the DPDF. Though the ledges support the 
general trends as the peaks, due to our limited 
knowledge of the true density functions, we decided not to use the information provided by 
the ledges.

\begin{figure}
\epsscale{0.85}
\plotone{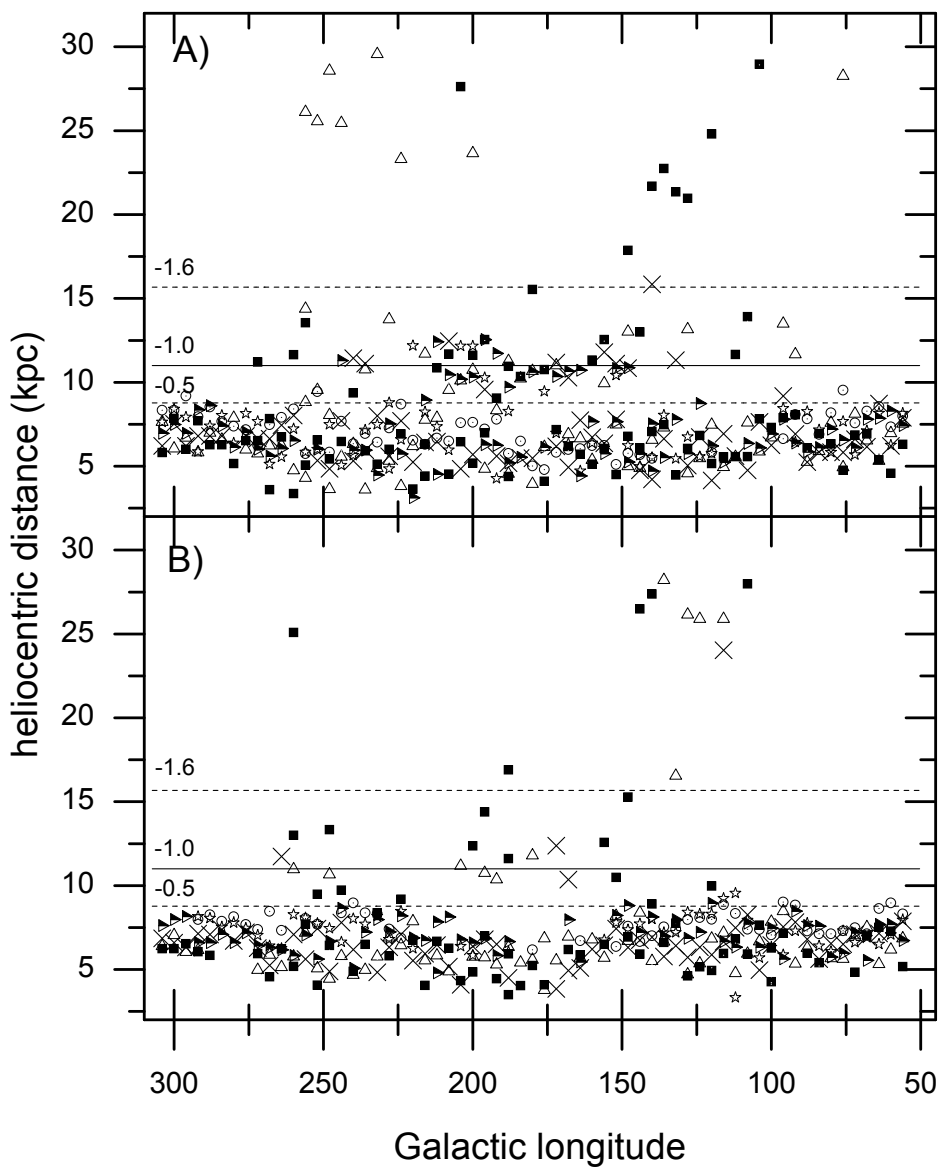}
\caption{Heliocentric distances of the main DPDF features found. The symbol indicates the 
Galactic latitude of the $4\degr\times 4\degr$ field for which a peak in the DPDF was found, 
as follows: solid squares, $36\degr$-$32\degr$; open triangles, 
$32\degr$-$28\degr$; crosses, $28\degr$-$24\degr$; half-filled right-pointing 
triangles, $24\degr$-$20\degr$; open stars, $20\degr$-$16\degr$; 
open circles, $16\degr$-$12\degr$. A solid and two dashed horizontal lines indicate 
the mean distance of the feature 
we identify as the Mon structure for varying average [Fe/H] in the MPDF ($-1.6$, $-1.0$ 
and $-0.5$ dex) as labelled in the plots. 
Panel A shows data for the North Galactic Hemisphere, panel B for the Southern Galactic Hemisphere. 
\label{peaksum}}
\end{figure}

Two main features can be seen in a summary of the detections of peaks (Fig. \ref{peaksum}). 
The large, closer strip comes from disk M giants, which
only appear in our sample beginning at $\sim 4$ kpc. 
The distance at which the bulk of disk stars appears is likely closer than shown,
because thin disk stars are likely to be more metal rich than the adopted mean metallicity in the MPDF.

The other main feature in Fig. \ref{peaksum} is the 
secondary strip around 11 kpc away from the Sun. 
This strip is very obvious in the Northern Hemisphere.  It corresponds to 
a preferred distance in the DPDF over a very large latitude range --- from 
$-36\degr < b < +36\degr$, 
although at much less significance in the southern detections due to substantially
more reddening in the Southern Hemisphere, which limited to a few the number of fields
we explored there.

Beyond 20 kpc from the Sun, a much farther group of DPDF features is apparent 
in both hemispheres.  While some could be related
to real structures, 
the majority of the distant peaks reflect one or two stars that create bumps in the 
DPDF, and at present we cannot assure that they are real.

\section{The Monoceros Stream}

We identify the structure defined by the second strip of DPDF peaks 
in Fig. \ref{peaksum} 
as the Mon halo stream (N02, Y03), a.k.a. the One Ring (I03). 
This is shown in Fig. \ref{ledgepeak}, where 
we give the M giant DPDF in $4\degr\times 4\degr$ fields centered at ($l,b$) 
for four fields where it was identified. 
The bimodal PDFs in each panel indicate the existence 
of two distinct stellar groups.

\begin{figure}
\epsscale{0.70}
\plotone{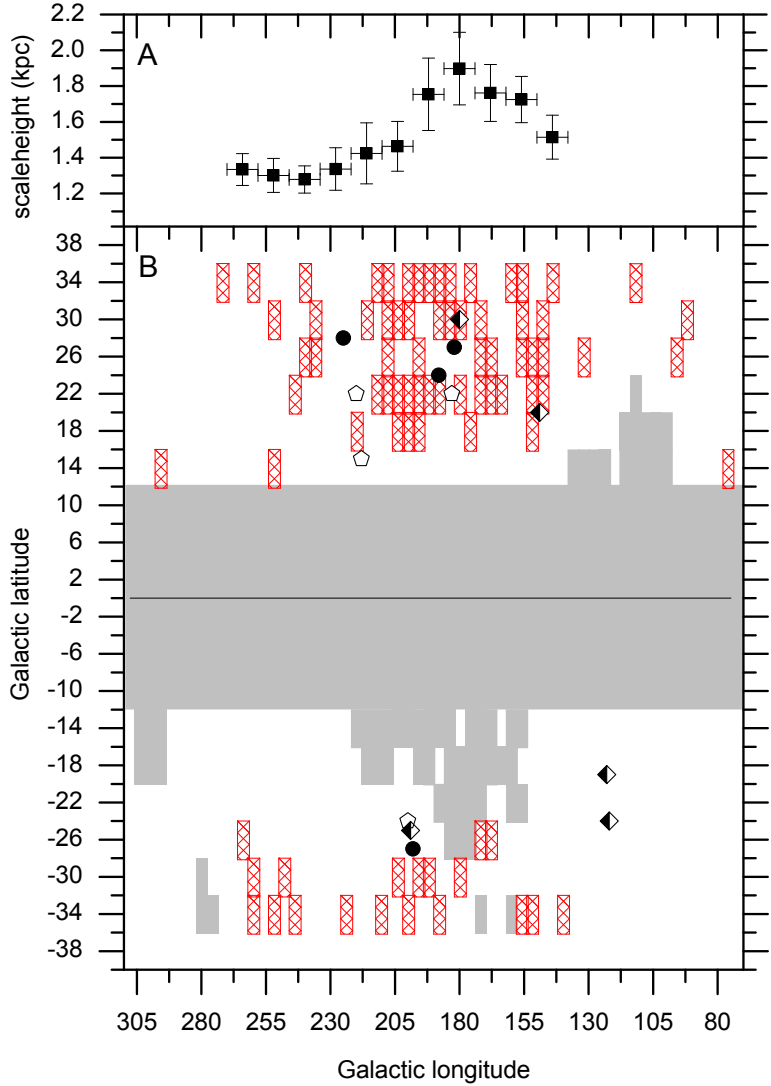}
\caption{Spatial and vertical distribution of the Mon Structure. 
Panel A shows the variations in the scaleheight of the Mon structure, 
calculated from all stars with most likely distances of 9-13 kpc in
the latitude range $+12\degr<b<+36\degr$. The structure seems to have a
large vertical extent 
towards $l\approx (180\pm 6)\degr$ and less towards $l\approx (240\pm 6)\degr$.
Panel B shows the likely signatures of the Mon structure already found in the literature. The large 
gray area shows fields where our analysis was hindered by reddening. Filled rectangles 
indicate fields where a DPDF peak at heliocentric 
distances 9-13 kpc is found. Other symbols give the central coordinates 
of fields where the Mon structure was found previously: pentagon, N02 fields; 
circles, Y03 fields; half-filled diamonds, I03 fields. The apparent edge in the structure for 
$|b|>36\degr$ is caused by the latitude cutoff of our analysis.
\label{ringtrace}}
\end{figure}

Our results confirm the existence of this peculiar Galactic structure in the outskirts of the 
disk. Although the structure was first announced as a possible tidal stream, I03 have 
pointed to its very large size, arguing that it might be more properly interpreted as a stellar ring, 
forming the very distant edge of the disk. 
Fig. \ref{ledgepeak} confirms
Ibata et al.'s findings in specific fields they explored, and Fig. \ref{ringtrace}b shows
that 2MASS traces the Mon feature over a large area of the sky.  In the data that we have 
analysed, the signal 
of the structure is strongest over $150\degr < l < 220\degr$.  From 
20-30\% of all M giants sampled in this area have most 
likely distances of 9-13 kpc. 
Closer to the Galactic plane, the structure is usually seen as 
a underlying peak (ledge) in the DPDF, due to the increasing dominance of disk M giants at 
smaller heliocentric distances.

In the North Hemisphere, peaks composing this structure have an 
average solar distance of 11.2 $\pm$ 0.11 kpc, in general agreement with 
the 8-11 kpc reported by N02, Y03 and I03. 
Previous authors 
report the Mon structure to be more distant in the South Hemisphere. 
Although our detected Southern Hemisphere DPDF features do
have a slightly larger distance (11.4 $\pm$ 0.24 kpc),  they do not show as wide a
north/south difference  as reported earlier ($\sim 2$ kpc). 
However, because we did not trace the structure as 
well in the south, the lack of substantial distance difference may not be significant.

\begin{figure}
\epsscale{0.80}
\plotone{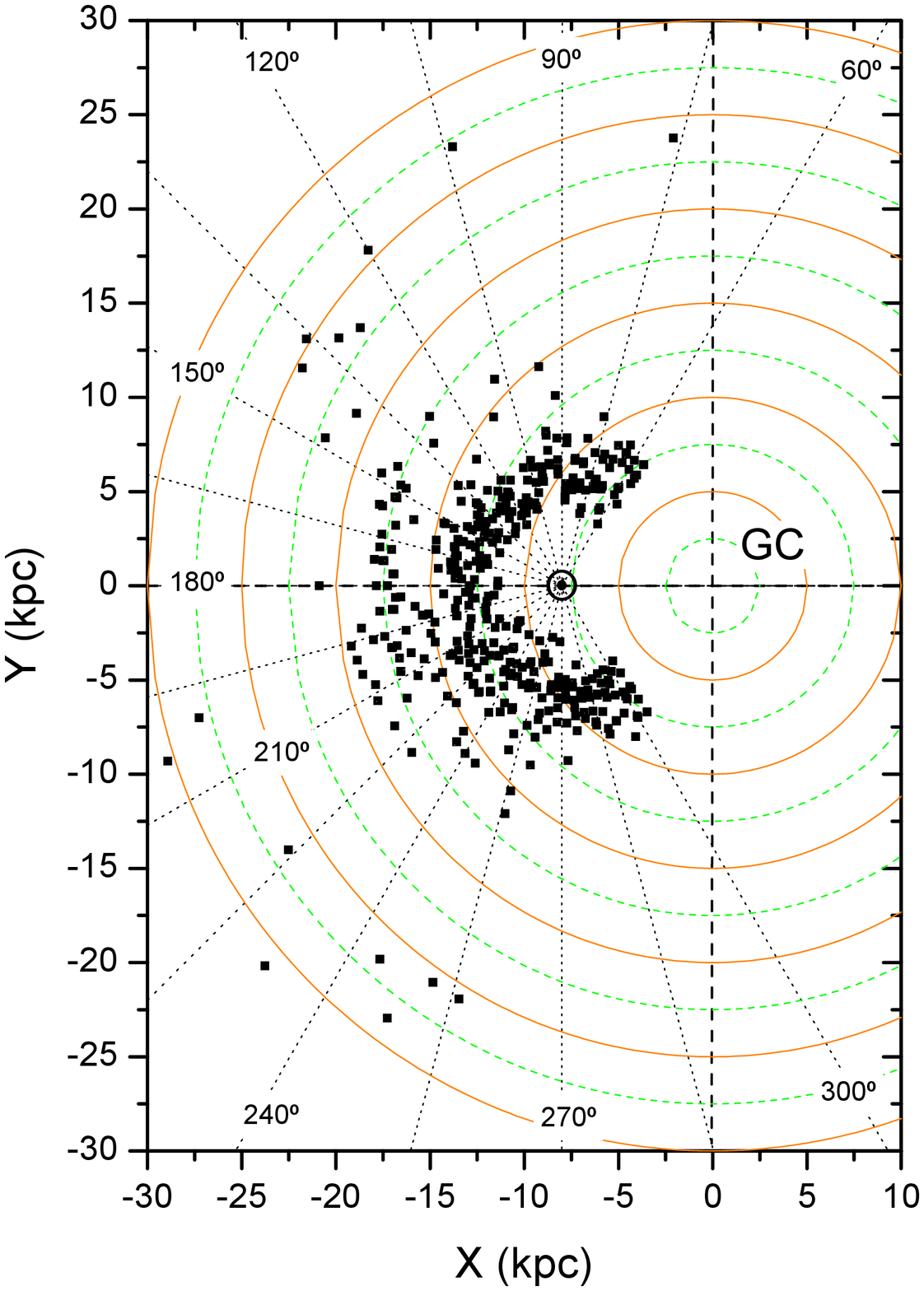}
\caption{Projection onto the Galactic plane of Northern hemisphere 
DPDF signatures along different lines of sight. The Mon structure is the 
thin feature $\sim 17.5\pm 0.5$ 
kpc away from the Galactic center. Curves of galactocentric isodistances are shown 
as a guide. Galactic longitudes are indicated by dotted lines. The Mon structure seems to 
be thicker between $195\degr < l < 240\degr$, merging with the disk somewhere around 
$l\approx 270\degr$.
\label{planeproj}}
\end{figure}

When projected on the Galactic plane, the structure is remarkably well seen 
(Fig.~\ref{planeproj}). 
As in Fig.~\ref{peaksum}, the thick belt encircling 
the solar position gives the distance where the bulk of disk stars are found.  
Broadening of this belt is partly a projection effect and partly due to 
the poor match of the adopted MPDF to disk stars.
The Mon structure appears as a generally 
thinner feature $\sim 17.5\pm 0.5$ kpc away from the Galactic center.
The derived mean {\it heliocentric} distance (projected onto the Galactic plane) 
is nearly uniform over a great
extent, from $l=130\degr$ to $210\degr$, 
but Fig.~\ref{planeproj}
conveys the impression that part of the structure may have significantly 
closer mean distance (perhaps even merging
with the disk) in the third Galactic quadrant compared to the second.  Based on
the scatter of peaks, the Mon feature appears to be deeper around 
$l = 200\degr$, but tapers to greater coherence for $l < 195\degr$, 
reminiscent of a dwarf galaxy with attached tidal tail.  

It is remarkable that although we have used an arbitrary MPDF, we have 
recovered stellar groups at the same position in the sky and nearly the same distance as 
those reported by N02, Y03 and I03. Tests show 
the thinness and distinctness of the Mon structure is remarkably robust to the adopted mean
[Fe/H] of the MPDF, though, of course, the distance scale varies.
The average distance of the structure is found to be approximated by 
\begin{equation}
\langle d\rangle = d_{0}\left(0.730 + 0.003 {\rm [Fe/H]} + 0.273 
{\rm [Fe/H]}^2\right) ,
\label{metvar}
\end{equation}
where $d_0$ is the average distance when $\langle{\rm [Fe/H]}\rangle = -1.00$. 
The effect of varying $\sigma$ is more complex.  With a broader MPDF, the DPDF both smears
out and shifts to smaller heliocentric distances.  This latter effect occurs because the metal-rich tail 
of the MPDF increases the probability 
of smaller distances for each star.  A large $\sigma$ would make nearly impossible the finding of 
an underlying blended peak, but the strongest features are still found:
e.g., in the $4\degr\times 4\degr$ 
field centered at $(l,b)=(180\degr,+30\degr)$ (corresponding to I03's WFS-0801 field) the 
Mon structure is found even with $\sigma$ as large as 1.00 dex. 
The distance of each peak found has its own uncertainty, which depends on the broadness 
of the peak. From the fields we have surveyed, we estimate roughly that this error is, 
at most, 2 kpc. Additionally, there is the uncertainty coming from our choice of the 
average [Fe/H] of the MPDF, which is nearly 2 kpc, calculated from error propagation 
with Eq.~\ref{metvar}. Adding these errors in quadrature, we have nearly a superior 
limit of 2.5 kpc for the average error in the distance of each peak. 

In \S1 we discussed previous evidence (M03) for the Mon feature in a sample
of 2MASS M giants even redder than used here.  The 
structure can be seen as a semi-detached blob centered at coordinates $(-20,-7)$ kpc
in Fig. 14e (and other panels of this figure) 
and $(175\degr, 11$ kpc) in the top panel of Fig. 10 in M03.  
This previous analysis uses a CMR derived from the Sagittarius core, expected
to be dominated by [Fe/H] $< -1$ stars, to assign a single 
distance to each star.  Though a different approach to that undertaken here, 
the Mon structure nevertheless shows nearly the same distance even when M03's
redder stars ($J-K_S > 1.10$) are considered.  
That the Mon feature is detected with even rather red M giants suggests the presence of
a population substantially more metal-rich than reported by Y03 --- a population
as poor as their [Fe/H] $= -1.6$ metallicity would have few tip giants in our sample
and even less in M03's reddest sample.   

Taking the simplistic approach that all of the sampled stars with 
most likely distance between 9 and 13 
kpc and $140\degr<l<270\degr$ 
are members of the structure, we calculate the mean scaleheight as $1.3 \pm 0.4$ kpc. This value 
is very similar to the one calculated by Y03 ($1.6\pm 0.5$ kpc), but nearly two times larger than 
the scaleheight estimated by I03.  However, because I03 adopted 8 kpc as the heliocentric
distance of the structure, 
this will have shortened their scaleheight estimate by $\sim$30\% relative to ours and Y03's. 
Y03 points out that the scaleheight could be as high as 3 kpc. 
Our estimate is preliminary, since at the moment we cannot separate a minor contamination 
by disk and halo stars.  
For the Northern Hemisphere fields explored, the Mon structure seems to have a larger 
scaleheight (Fig.~\ref{ringtrace}a) towards $l\approx (180\pm 6)\degr$ and a smaller one
towards $l\approx (240\pm 6)\degr$.   Accounting also for the trend of lower
scaleheight with decreasing $l$ in the second quadrant, the collective density
distribution bears the hallmarks of a larger core structure near $l = 200\degr$ with 
thinner, more coherent ``tail" features in either direction.     

Fig.~\ref{planeproj}, which gives 
the impression that the structure is arcing into the Galactic disk and merging with
it beyond $l \approx 270\degr$,
suggests that the latter explanation is more likely.  Moreover, in spite of being very large, 
Fig.~\ref{planeproj} suggests that the structure also disappears at $l < 90\degr$.
On the other hand, inside $-90\degr < l < +90\degr$ the 2MASS M giants
begin to sample the inner Galaxy, and the disappearance of 
signatures of the Mon structure might also be explained by the preponderance 
of disk giants dominating the DPDF, making Mon peaks progressively
more difficult to see.
In any case, the present analysis suggests that the Mon structure is not
a uniform ring in terms of distance, depth, vertical extent or density.

\citet{helmi} have analysed dynamical simulations of mergers 
for the creation of ring-like structures similar to Mon and
discuss two possible scenarios for its origin that give rise to very different
spatial distributions:  (1) Recent accretion of a satellite galaxy in nearly coplanar 
orbit with the Galactic disk would form a transient arc of non-uniform
Galacticentric distance
The feature would have a large 
number of stars and could obviously present a varying vertical distribution with 
respect to the Galactic equator.  (2) An ancient minor 
merger would presently be seen as concentric shells around the Galaxy, 
symetrically distributed with respect to the Galactic plane. 
Our results more strongly resemble the Helmi et al. models of the first situation:
The Mon feature looks like a planar arc with a central
core, greater coherence away from that core, and varying Galactocentric distance.  Combined
with the presence of relatively metal-enriched stars of the type expected in 
younger stellar systems, a tidally disrupted dwarf spheroidal scenario, similar
to the Sgr system, seems most likely. 

Although we have adopted the name Monoceros to this Galactic Anticenter structure, 
in order to be consistent with previous authors, we have found that the structure spans 
a large range of constellations. Since its center is yet unclear, other names 
may need to be considered after more data become available.


\begin{thebibliography}{}


\bibitem[Dutra \& Bica(2001)]{dutrabica} Dutra, C.~M.~\& Bica, 
E.\ 2001, \aap, 376, 434 

\bibitem[Dutra et al.(2003)]{dutra} 
Dutra, C.~M., Santiago, B.~X., Bica, E.~L.~D., \& Barbuy, B.\ 2003, \mnras, 
338, 253 

\bibitem[Helmi et al.(2003)]{helmi} Helmi, A., Navarro, J. F., Meza, A., 
Steinmetz, M., \& Eke, V. R. 2003, preprint (astro-ph/0303305)

\bibitem[Ibata et al.(2003)]{ibata} Ibata, R. A., Irwin, M. J., Lewis, 
G. F., Ferguson, A. M. N., Tanvir, N. 2003, accepted by MNRAS (astro-ph/0301067)

\bibitem[Ivanov \& Borissova(2002)]{ivanov} Ivanov, V. D., Borissova, J. 2002, 
\aap, 390, 937

\bibitem[Ivanov et al.(2002)]{ivanovetal} Ivanov, V.~D., 
Borissova, J., Pessev, P., Ivanov, G.~R., \& Kurtev, R.\ 2002, \aap, 394, 
L1 

\bibitem[Majewski et al.(2003)]{maj03} Majewski, S. R., Skrutskie, M. F., Weinberg, M. D., 
Ostheimer, J. C. 2003, submitted (astro-ph/0304198)

\bibitem[Newberg et al.(2002)]{newberg} Newberg, H. J., Yanny, B., Rockosi, C. M., et al. 
2002, \apj, 569, 245

\bibitem[Press et al.(1989)]{press}
Press, W. H., Teukolsky, S. A., Vetterling, W. T., Flannery, B. P., {\it
Numerical Recipes in Pascal: The Art of Scientific Computing}, Cambridge
Univ. Press, New York, 1989, p. 222

\bibitem[Skrutskie et al.(2001)]{skrutskie} 
Skrutskie, M.~F., Reber, T.~J., Murphy, N.~W., \& Weinberg, M.~D.\ 2001, 
American Astronomical Society Meeting, 199,  

\bibitem[Weinberg \& Nikolaev(2001)]{weinberg} Weinberg, 
M.~D.~\& Nikolaev, S.\ 2001, \apj, 548, 712 

\bibitem[Yanny et al.(2003)]{yanny} Yanny, B., Newberg, H. J., Grebel, E. K. 2003, et al. 
2003, \apj, 588, in press 

\bibitem[Yoss, Neese \& Hartkopf(1987)]{yoss} Yoss, K. M., Neese, C. L., and Hartkopf, 
     W. I. 1987, \aj, 94, 1600


\end{thebibliography}
\end{document}